\documentclass[
amsmath, %
floatfix, %
twocolumn, %
reprint, %
superscriptaddress, %
prb, %
aps, %
citeautoscript, %
final, %
]{revtex4-1} %arxiv OFF
\renewcommand{\thispagestyle}[1]{}
\usepackage[]{newtxtext} %arxiv OFF
\usepackage[subscriptcorrection,nosymbolsc,smallerops,bigdelims]{newtxmath} %arxiv OFF
\DeclareMathAlphabet{\mathcal}{OMS}{cmsy}{m}{n} %arxiv OFF
\DeclareMathAlphabet{\mathbcal}{OMS}{cmsy}{b}{n} %arxiv OFF
\usepackage{bm}

\usepackage[utf8]{inputenc}
\usepackage[T1]{fontenc}

\usepackage[]{graphicx}
\usepackage{latexsym}

\usepackage{wasysym}

\usepackage{color}
\usepackage{mathtools}
\usepackage[]{xcolor}
\usepackage{bm}
\usepackage{siunitx}
\usepackage{hyperref}
\hypersetup{
	colorlinks,
	linkcolor={blue!90!black},
	citecolor={blue!90!black},
	urlcolor=	{blue!90!black}
}
\let\oldeqref\eqref
\renewcommand*{\eqref}[1]{%
	Eq.~\hyperref[eq:#1]{\oldeqref{eq:#1}}%
}

\usepackage{floatrow}
\mathchardef\mhyphen="2D

\DeclarePairedDelimiter\abs{\lvert}{\rvert}

\DeclarePairedDelimiterX{\comm}[2]{\lbrack}{\rbrack}{#1, #2}

\DeclarePairedDelimiterX{\braket}[2]{\langle}{\rangle}{#1\delimsize\vert #2}
\DeclarePairedDelimiterX{\ketbra}[2]{\rvert}{\lvert}{#1 \delimsize\rangle\!\delimsize\langle #2}
\DeclarePairedDelimiterX{\matrixel}[3]{\langle}{\rangle}{#1 \delimsize\vert #2 \delimsize\vert #3}

\newcommand{\subfigref}[2]{Fig.~\hyperref[fig:#1]{\ref*{fig:#1}(#2)}}
\newcommand{\subfigsref}[3]{Figs.~\hyperref[fig:#1]{\ref*{fig:#1}(#2)}-\hyperref[fig:#1]{\ref*{fig:#1}(#3)}}

\usepackage[low-sup]{subdepth}
\usepackage{ragged2e}

\definecolor{cbred}{HTML}{e31a1c}
\definecolor{cbgreen}{HTML}{33a02c}
\definecolor{cbblue}{HTML}{176aa7}
\definecolor{cborange}{HTML}{ff7f00}
\definecolor{cbviolet}{HTML}{6a3d9a}
\definecolor{cbbrown}{HTML}{b15928}

\definecolor{cblred}{HTML}{fb9a99}
\definecolor{cblgreen}{HTML}{b2df8a}
\definecolor{cblblue}{HTML}{a6cee3}
\definecolor{cblorange}{HTML}{fdbf6f}
\definecolor{cblviolet}{HTML}{cab2d6}
\definecolor{cblbrown}{HTML}{ffff99}

\usepackage[activate={true,nocompatibility},final,tracking=alltext,kerning=true,spacing=true,protrusion=true,factor=1100,stretch=4,shrink=6,selected=true ,letterspace=-0]{microtype}

\usepackage{orcidlink}

\begin{document}

\title{Disorder-resilient transport through dopant arrays in silicon}

\author{Micha\l\ Gawe{\l}czyk\orcidlink{0000-0003-2299-140X}}
\email{michal.gawelczyk@pwr.edu.pl}
\affiliation{Institute of Physics, Faculty of Physics, Astronomy and Informatics, Nicolaus Copernicus University in Toru\'n, Grudzi\k{a}dzka 5, 87-100 Toru\'n, Poland}
\affiliation{Institute of Theoretical Physics, Wroc\l{}aw University of Science and Technology, 50-370 Wroc\l{}aw, Poland}

\author{Garnett W. Bryant\orcidlink{0000-0002-2232-0545}}
\affiliation{Nanoscale Device Characterization Division, Joint Quantum Institute, National Institute of Standards and Technology, Gaithersburg, Maryland 20899-8423, USA}
\affiliation{University of Maryland, College Park, Maryland 20742, USA}

\author{Micha{\l} Zieli\'nski\orcidlink{0000-0002-7239-2504}}
\affiliation{Institute of Physics, Faculty of Physics, Astronomy and Informatics, Nicolaus Copernicus University in Toru\'n, Grudzi\k{a}dzka 5, 87-100 Toru\'n, Poland}

\begin{abstract}
Chains and arrays of phosphorus donors in silicon have recently been used to demonstrate dopant-based quantum simulators. The dopant disorder present in fabricated devices must be accounted for. Here, we theoretically study transport through disordered donor-based $3\times 3$ arrays that model recent experimental results. We employ a theory that combines the exact diagonalization of an extended Hubbard model of the array with a non-equilibrium Green’s function formalism to model transport in interacting systems. We show that current flow through the array and features of measured stability diagrams are highly resilient to disorder. We interpret this as an emergence of uncomplicated behavior in the multi-electron system dominated by strong correlations, regardless of array filling, where the current follows the shortest paths between source and drain sites that avoid possible obstacles. The reference $3\times 3$ array has transport properties very similar to three parallel 3-site chains coupled only by interchain Coulomb interaction, which indicates a challenge in characterizing such devices.
\end{abstract}

\maketitle
	
\section*{Introduction}
    Recently there has been much interest in using semiconductor quantum dot arrays and precisely placed dopants in Si as sites to carry out analog quantum simulations (AQS) of extended Hubbard models. Nagaoka ferromagnetism was demonstrated experimentally for the first time with a $2\times 2$ array of gated semiconductor quantum dots \cite{DehollainNature2020}, more than fifty years after the predictions of Nagaoka \cite{NagaokaPhysRev1966}. The Su-Schrieffer-Heeger model \cite{SuPRL1979} was simulated using a one-dimensional chain of phosphorous dopant quantum dots in Si \cite{KiczynskiNature2022}. Precise placement of the dopants has been done with scanning tunneling microscope-based lithography \cite{WyrickAFM2019} and used to fabricate $3\times 3$ dopant arrays for AQS \cite{WangNatureComm2022}. The $3\times 3$ array with 3 rows, each with 3 dopant sites, connecting the source to the drain is the smallest two-dimensional array with at least one column of sites not directly coupled to the source or the drain by tunneling. The study of the $3\times 3$ array is the starting point for the simulation of extended two-dimensional systems with internal sites and edge sites not directly coupled to the leads.  

    In their study of $3\times 3$ dopant arrays, Wang \textit{et al.} \cite{WangNatureComm2022} identified a delocalized-localized transition when the intersite separation increased and the tunneling between sites decreased. In the delocalized regime, charge stability diagrams displayed charge transition boundaries all with the same slope, indicating the delocalization of the array states into the states of a single large quantum dot, corresponding to a metallic state in an extended system. At large site separation, three slopes for the charge transition lines were seen after the transition to the localized phase. Each slope corresponded to the current along one of the three rows of sites between the source and drain, with the side-gate bias on each row experiencing a different dependence on the two side gates due to the different lever arms between each row and each side gate. 

    In AQS with gate-defined quantum dots, the gates can be used to tune the quantum dots and the coupling between the dots. This tunability is an important advantage for gated semiconductor AQSs. One-dimensional structures, (\textit{i.e.}, $1\times N$ chains of dots) and two-dimensional structures limited to two parallel chains, (\textit{i.e.}, $2\times N$ chains) can be fabricated with gated quantum dots. However, other geometries will be much harder to realize with gated structures. In particular, two-dimensional structures with both internal sites and edge sites should be more easily realized by dopant placement. The $3\times 3$ dopant arrays studied by Wang \textit{et al.} \cite{WangNatureComm2022} represent the first step in this direction. With dopant arrays, tunability is achieved by making arrays with different spacing between sites or with different dopant configurations at the sites. Perfect placement of individual dopants will be needed for the controlled fabrication of complex array geometries.

    Perfect dopant placement may soon be realized \cite{WyrickACSNano2022}. However, disorder due to fluctuations in dopant position or dopant number at a site still must be accounted for in dopant devices currently being used for AQS. In this paper, we show theoretically that current flow through a $3\times 3$ dopant array is highly resilient to disorder. We observe current flow characterized by stability and Coulomb-blockade diagrams, which indicates that electrons just move around local obstructions created by the disorder. Consequently, transport measurements can be a useful probe of AQS even in the presence of disorder. 

    \begin{figure*}[tb]
        \includegraphics[width=0.98\linewidth]{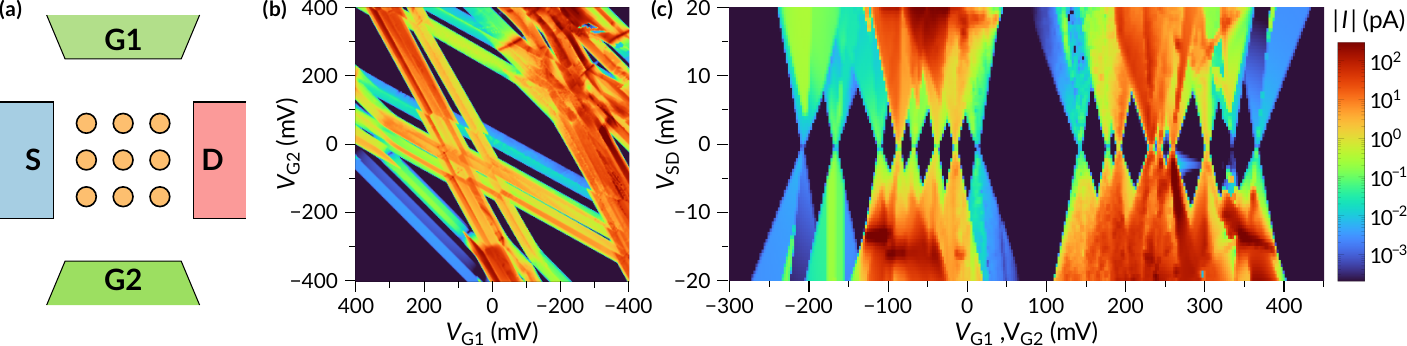} %
        \caption{\label{fig:exp} Simulation of the experimental device. (a) Schematic diagram of the $3\time 3$ system with source (S), drain (D), and plunger gates (G1, G2) marked. (b) Stability diagram showing the total source-drain current through the device (color scale) for a fixed bias voltage $V_{\mathrm{SD}} = 8$~mV as a function of plunger gate potentials $V_{\mathrm{G_1}}$ and $V_{\mathrm{G_2}}$. (c) Coulomb blockade diagram showing the total source-drain current through the device (color scale) as a function of the plunger-gate potential $V_{\mathrm{G_1}}=V_{\mathrm{G_2}}$ and bias voltage $V_{\mathrm{SD}}$ and. Model parameters are taken from Ref.~\onlinecite{WangNatureComm2022} to reflect the actual experimental device.} %
    \end{figure*}

    We use the extended Hubbard model to study dopant arrays. The extension is needed as the sites are not electrically neutral and are close enough that the Coulomb and exchange interactions between the sites play a significant role. To calculate the current, we must deal with an open system coupled to the leads. For this, we follow the non-equilibrium Green's functions formalism. The system is strongly correlated, \textit{i.e.}, Coulomb interactions are much stronger than hopping, and mean-field approximations are insufficient. Thus, we resort to constructing Green's functions by directly finding many-electron eigenstates. This approach based on combining exact diagonalization with non-equilibrium Green's functions allows us to calculate terminal currents in the arrays and other observables characterizing the system. Typically, by resorting to Green's functions, one loses information on the underlying eigenstates. Our methodology lets us preserve this knowledge and back-trace states responsible for features of interest. This gives us a way to gain more information about the transport processes than can be obtained directly from transport measurements. Thus, we can, \textit{e.g.}, characterize the many-body configurations that contribute to the current or determine current magnitudes for different channels. We discuss these effects to better understand transport studies of experimentally realized arrays \cite{WangNatureComm2022}.

\section*{Results and discussion}\label{sec:results}

\subsection*{The device and its model}

    \begin{figure*}[tb]
        \includegraphics[width=\linewidth]{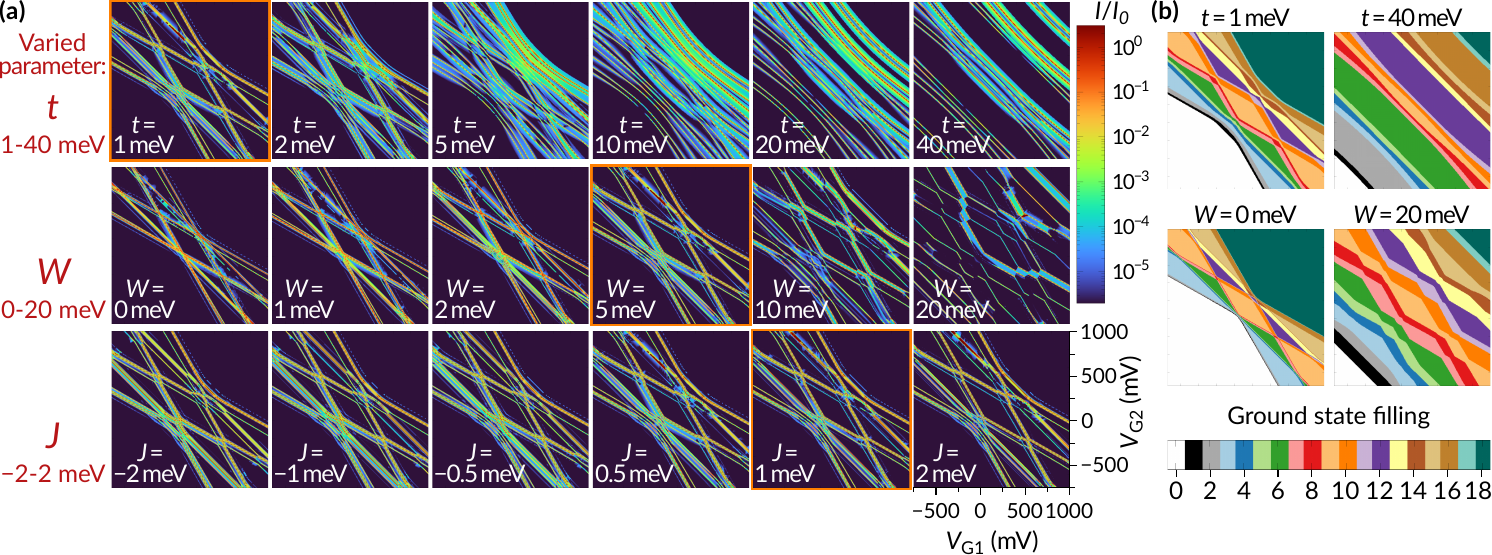} %
        \caption{\label{fig:params} Impact of model parameters on transport properties of the system. (a) Stability diagrams for $V_{\mathrm{G2}}$ and $V_{\mathrm{G1}}$ varied in the range of $-750$--1000~mV (as shown in the lower right panel). In each row, one of the parameters is varied, while the other parameters are fixed at values representing an ideal reference device: $t=1$~meV, $U=44$~meV, $W=5$~meV, $J=1$~meV (framed in each row). (b) Ground state filling (electron number) of the array shown for the first and last cases of varying $t$ and $W$.} %
    \end{figure*}
    
    To keep our study closely connected to experimentally realized AQSs, we take the main system studied in Ref.~\onlinecite{WangNatureComm2022} as a reference. It consists of a $\approx 10.7$-nm-spaced $3\times3$ array of few-dopant sites tunnel coupled to the source and drain gates and capacitively coupled to two side gates. The last two serve as plunger gates or enable the application of a transverse electric field. Together with two other arrays with smaller site spacings, which provide a smaller ratio of Coulomb repulsion to hopping, $U/t$, Wang \textit{et al.} demonstrated the transition to the metallic regime. We show the schematic view of the system in Fig.~\ref{fig:exp}(a).
	
    We model the system with the extended Hubbard model that accounts for the hopping of electrons between the sites ($t$), on-site ($U$), and long-range ($W$) Coulomb repulsion of electrons, as well as their attraction to positively charged sites and nearest-neighbor electron-electron exchange ($J$). For a gated system, we additionally evaluate the lever arms between the leads and each of the sites \cite{WangNatureComm2022}, which gives us site-dependent chemical potentials for a given configuration of lead potentials. To study transport properties, we employ the powerful non-equilibrium Green's functions (NEGF) formalism \cite{KeldyshZETF96,MahanBook2000,HaugBOOK2008}. As the system is strongly correlated, \textit{i.e.}, Coulomb interactions are much stronger than hopping, and mean-field approximations are insufficient. For this reason, we construct Green's functions by directly finding many-electron eigenstates. Our combination of exact diagonalization with the NEGF method is thus nonperturbative in contrast to the standard NEGF approach. Electron states in the source and drain leads are described via a proper self-energy and tunnel coupled to the array. As this coupling has nonvanishing terms only between states in the array and the lead, its inclusion is also nonperturbative with the only approximation assuming leads are screened (metallic) and remain in equilibrium. As a result, we get a treatment that is in principle exact within the model's assumptions at the cost of solving the Hubbard model for all charge-spin sectors of Fock space at each set of system parameters studied. After including the coupling to the leads, we can calculate the terminal current in the system. We give the details of the model and our approach in the Methods section.

    We begin by presenting in Fig.~\ref{fig:exp} the results we obtained for the widely spaced $3\times 3$ array from Ref.~\onlinecite{WangNatureComm2022}. We use the site-dependent model parameters as assessed by the authors. This site dependence is due to disorder, as the number of donors and their arrangement within a site may vary. Later in the paper, we systematically study systems without and with decomposed disorder as well. 

    Fig.~\ref{fig:exp}(b) shows the stability diagram obtained by separately varying the two plunger-gate potentials at a fixed low source-drain bias of $V_{\mathrm{SD}}$ = $8$~mV. We observe charge transition lines characterized by nonzero currents in the diagram. The slope of the lines has approximately three values, which correspond to the three positions of sites with respect to plunger gates (for each row of sites). The middle row of sites is roughly equally distant from both leads, resulting in antidiagonal lines, while the other two slopes, steeper and less steep than the antidiagonal, correspond to sites lying closer to one of the leads. As the actual device has a disorder, we observe little degeneracy and some deviations from the expected tripartite pattern. We notice the evident wide separation of the charge stability diagram into sections below and above half-filling with the Mott gap and the Coulomb-blockaded diamond-shaped region in the middle.

    Notably, currents are generally higher in the high-filling regime. This observation is surprising at first glance. A weaker current might be expected above half-filling due to interactions blocking the movement of electrons. We explain the higher currents above half-filling by realizing that populating the system with electrons smoothes the potential landscape, partially limiting the influence of disorder. This intuition may be unreliable in a highly correlated system, but it seems justified when crossing the specific half-filling point. Experimental devices also manifested this effect \cite{WangNatureComm2022}. Another noticeable difference between the two parts of the diagram is the presence of "bridges" at line anticrossings corresponding to tunneling resonances \cite{VaartPRL1995,WeberNN2014}. These are most evidently visible around the diagram's $(V_{\mathrm{G_1}}=50~\mathrm{meV}, V_{\mathrm{G_2}}=300~\mathrm{meV})$, and $(V_{\mathrm{G_1}}=350~\mathrm{meV}, V_{\mathrm{G_2}}=300~\mathrm{meV})$ points and were also observed for the experimental device and attributed to hybridization (due to tunneling) of many-body states with distinct charge distributions \cite{WangNatureComm2022}.
    Reproducing these two effects shows that the model correctly captures the impact of delocalization and correlation in the system.

    In panel (c) of Fig.~\ref{fig:exp}, we present the complementary Coulomb-blockade diagram showing the current in the system for various source-drain biases as the potentials of plunger gates are tied together and swept. These results correspond to changing bias along the diagonal of the stability diagram, with more electrons occupying the array with increasing chemical potential. We observe that each conduction resonance typically gets widened at a higher bias into an hourglass-shaped feature. For the current to flow, a resonance must lie energetically in the bias window, which is widened proportionally to the applied voltage. The features form two groups, below and above half-filling, with a pronounced Coulomb blockade diamond in the middle. Here also, \textit{e.g.}, at $V_{\mathrm{G_{1,2}}}=-120~\mathrm{meV}$ and $V_{\mathrm{G_{1,2}}}=350~\mathrm{meV}$, we observe additional horizontal enhancements corresponding to tunnel coupling and complementary to those we noticed in the stability diagram. Finally, we also see that the summation of currents is not always constructive when two features overlap. We notice symptoms of partially destructive interference, \textit{e.g.}, at $V_{\mathrm{G_{1,2}}}=-145~\mathrm{meV}$ for negative biases from $V_{\mathrm{SD}}$ = $-14$~mV.

    \begin{figure*}[tb]
    	\includegraphics[width=0.8\linewidth]{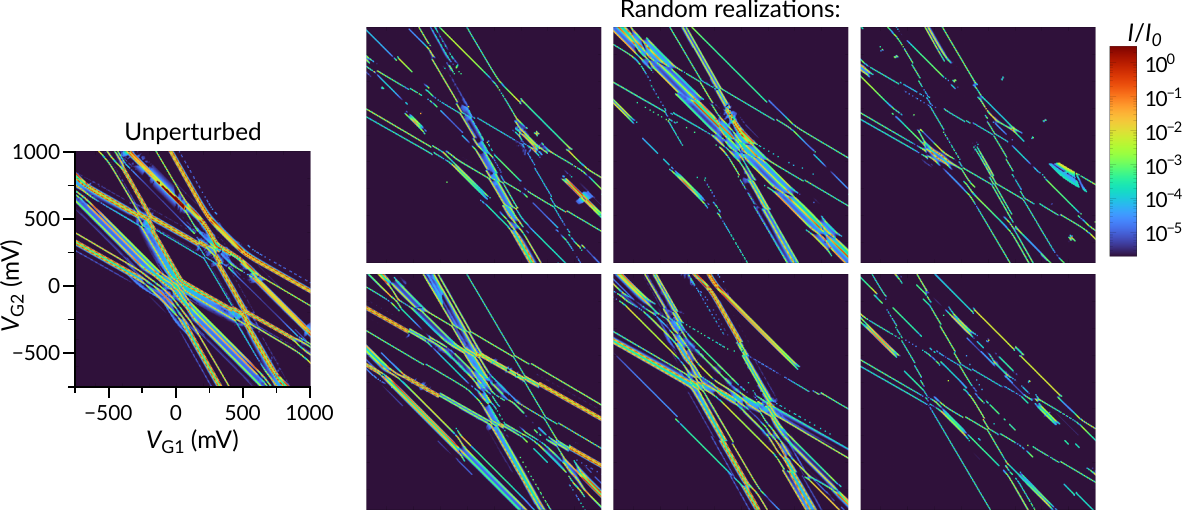} %
    	\caption{\label{fig:disorder} Impact of disorder. Each of the six panels on the right shows a stability diagram for a different realization of site disorder in the lattice. For each of the sites the on-site energy $E_{\mathrm{B}}$ is varied by a value drawn from the set $\{-20, 0, 20 \}$~meV that simulates the $\pm1$ uncertainty for the number of donors forming the site. On the left, the diagram for the unperturbed system is shown for comparison.}  %
    \end{figure*}
    
    \begin{figure*}[tb]
    	\includegraphics[width=0.76\linewidth]{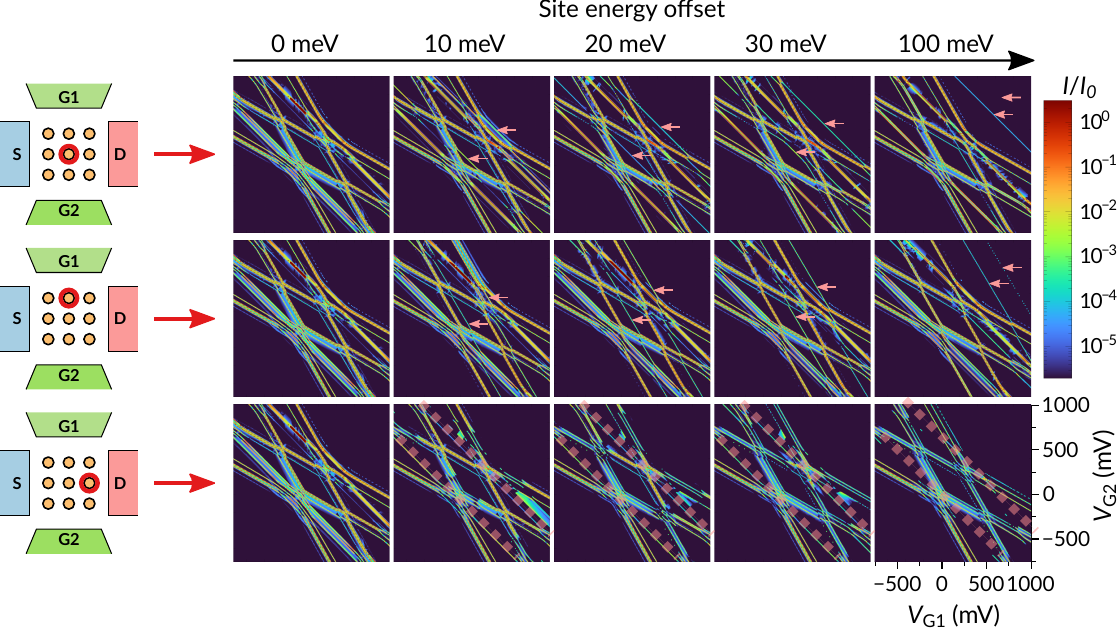} %
    	\caption{\label{fig:offsets-one} Disorder decomposition into single-site energy offsets. Each row shows the evolution of the stability diagram as the on-site energy for the given site is changed from $0$ to $100$~meV with respect to the rest of the lattice. In the first two rows, pale red arrows show the diagram lines that are shifted with the increasing offset; in the bottom row, thick dashed pale-red lines show positions of lines that vanish due to the offset.} %
    \end{figure*}
    
    \begin{figure*}[tb]
    	\includegraphics[width=0.65\linewidth]{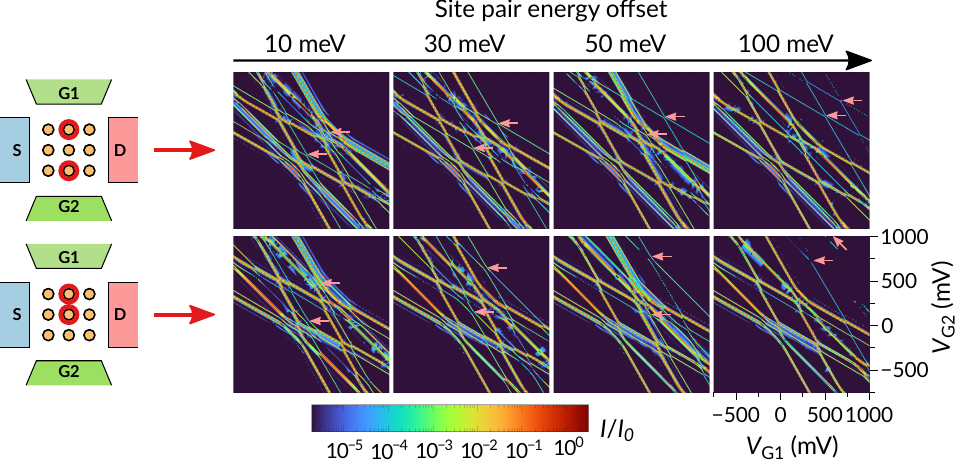} %
    	\caption{\label{fig:offsets-two} Current through a single-site bottleneck. Each row shows the evolution of the stability diagram as the on-site energy for the given pair of sites is changed from $0$ to $100$~meV with respect to the rest of the lattice.} %
    \end{figure*}

\subsection*{Model parameters and the transition to the metallic state}
    In the following, we first consider a $3\times 3$ array without disorder to establish the response of an idealized, perfectly ordered array. Then, we systematically study the consequences of deviations from this idealized reference. We set the array parameters to be: hopping $t = 1$~meV, on-site repulsion $U = 44$~meV, long-range Coulomb $W = 5$~meV, and nearest-neighbor exchange of $J = 1$~meV. In particular, none of the parameters is now position-dependent. It should be noted, however, that the array potential is still not homogeneous because the array is finite and the Coulomb potential due to attraction to positively charged sites is largest at the middle site in the array \cite{LePRB2017a}.

    We focus on the idealized array to understand features observed in the realistic array. We vary model parameters independently. We show the results in Fig.~\ref{fig:params}(a), where each row of stability diagrams corresponds to changing a single parameter, while the other parameters are kept fixed at values corresponding to the parameters of the ideal, reference array. In each row, the panel framed in orange is for the ideal reference array. Additionally, Fig.~\ref{fig:params}(b) shows the ground state filling for selected diagrams corresponding to the lowest and highest studied values of $t$ and $W$.
 
    In the first row of Fig.~\ref{fig:params}(a), we show the transition to the metallic (delocalized) regime when hopping $t$ increases and becomes comparable with Coulomb repulsion. As the eigenstates become more delocalized due to increasing hopping, we observe a widening of the line anticrossings, and finally, the three slopes initially present in the diagram merge into a single antidiagonal slope. This simplification of the charge stability diagram reflects that a state delocalized over the entire array is equally distant to both plunger gates, and the device effectively behaves as a single quantum dot (site). Importantly, in this regime, the Coulomb-blockade diamonds are almost completely closed.

    Next, we vary the strength of the long-range Coulomb interaction of electrons in the array. The results are shown in the second row of panels in Fig.~\ref{fig:params}. While the largest isolating diamond-shaped areas correspond to the on-site repulsion, the long-range interaction introduces additional splittings of the observed lines. These splittings correspond to lifting the degeneracy present at $W = 0$ and resulting from electrons' freedom to choose a site as long as it is not already occupied.

    Finally, we check the impact of nearest-neighbor exchange energy $J$, an energy penalty for the parallel alignment of nearest-neighbor spins. We vary $J$ in a wide range of values reaching far beyond both sides of the ferro-antiferromagnetic transition. While this interaction is essential for the magnetic properties of the array, the third row of panels in Fig.~\ref{fig:params} shows $J$ has minimal effect on the probed charge transport properties. The only noticeable difference $J$ introduces is the minor changes in the splittings caused by long-range Coulomb interaction. Thus, to quantify $J$, critical for quantum-information applications of impurity systems \cite{LevyPRL2002,ScarolaPRA2005,HeNature2019}, a more sophisticated experimental methodology will be required, like the one used in Ref.~\onlinecite{DehollainNature2020} for a system of gate-defined quantum dots. While not visible in the transport properties, we identify the paramagnetic-ferromagnetic transition in the reference system at approximately $J = -0.1$~meV. 

\subsection*{Impact of disorder}

    Although the technology of positioning phosphorus impurities in silicon has achieved near atom-scale precision, a certain amount of disorder is still inevitable. Disorder due to small inaccuracies in the positions of the array sites can translate into large hopping and exchange integral fluctuations \cite{KoillerPRL2001,KoillerPRB2002,WellardPRB2003,PicaPRB2014,GamblePRB2015,MohiyaddinPRB2016,SongAPL2016}. These fluctuations were due to oscillations in the dependence on the donor separation resulting from valley interference \cite{SalfiNC2014,VoisinNC2020,TankasalaPRB2022}. However, this problem concerns systems in which single impurities comprise the sites. In fact, systems of this kind are not yet typically manufactured. Creating sites containing several donors is more feasible and advantageous for suppressing those oscillations \cite{WangNJLQI2016}. In this situation, minor inaccuracies in the positions of sites become less relevant, and the randomness of on-site energy related to the dispersion in the number of donors forming a site is the dominant source of disorder.

    Calculations of the electron binding energy in few-dopant quantum dots \cite{WeberNN2014} showed that the on-site energy for a singly, positively-charged site depends linearly on the number of phosphorus atoms forming the site, with an increase in binding of approximately $20$~meV per atom. Based on the characterization of the actual device in Ref.~\onlinecite{WangNatureComm2022}, we infer that a good description of this disorder is to adopt a discrete tri-modal distribution of on-site energies: the base one and $\pm20$~meV, \textit{i.e.}, the set $\{-20, 0, 20\}$~meV, each with equal probability. We present corresponding results showing six realizations of such a disorder in Fig.~\ref{fig:disorder}. While individual lines are randomly shifted and current values are generally lower, the pattern is largely retained. We can still recognize the contributions (slopes) associated with each row in the array. Noticeably, for some realizations, all lines with a given slope are stronger than the lines with other slopes, indicating a common source affecting the entire conduction "channel". This interpretation will be further supported by results presented later.

    To understand the observed effects of disorder more easily, we decompose the disorder into individual contributions. To do this, we return to the reference array and check the effects of changing the on-site energy at only one site at a time. The corresponding results are presented in Fig.~\ref{fig:offsets-one}, where each line shows a change in the diagram due to the gradual energy shifting on the indicated site. The first two rows concern the middle sites not connected to the leads. Although the electrons in the array are strongly correlated, the effect of single-site disorder is surprisingly simple. A single pair of lines (related to single and double site occupation) with a slope corresponding to the given site's row moves as the site energy changes. After it is pushed outside the diagram, corresponding to the exclusion of a given site from transport, the diagram is qualitatively intact, just missing a single pair of lines.
	
    The situation is different when the single-site disorder concerns a site coupled to one of the leads (see the bottom row in Fig.~\ref{fig:offsets-one}). In such a case, all lines with a given slope decay, as exemplified by the disappearance of antidiagonal lines in the plotted case. This result points to the crucial role of terminal sites that connect to the source and drain and suggests that groups of lines with a given slope are, in fact, associated with spatial conduction channels determined by which terminal sites play an essential role.

\begin{figure*}[tb]
	\includegraphics[width=0.83\linewidth]{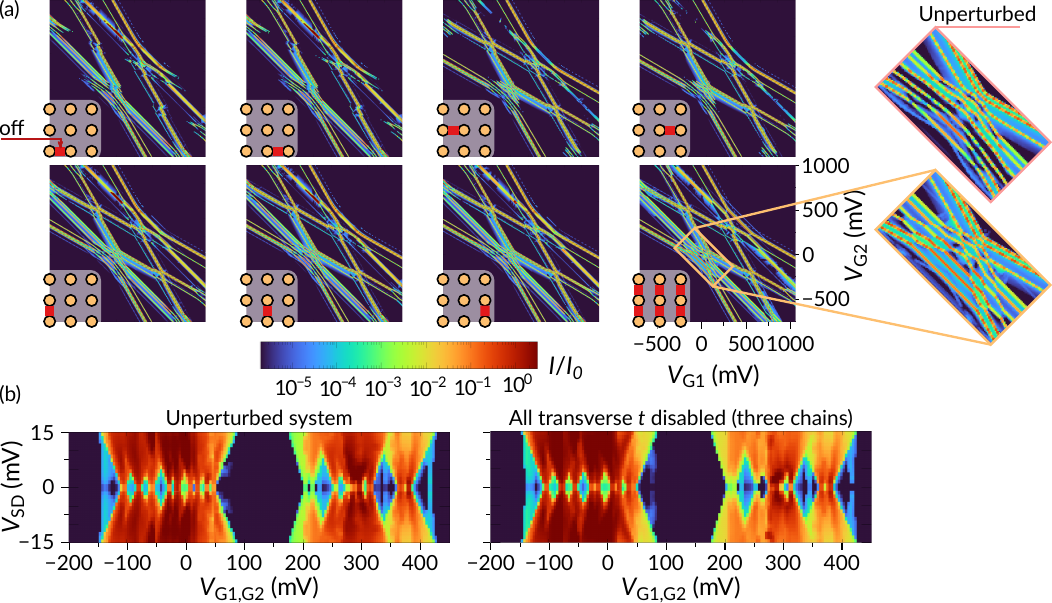} %
	\caption{\label{fig:hopping-cut} Impact of disabled hopping. Part (a) shows stability diagrams for all nonequivalent cases of a single disabled hopping in the array and for the case of all transverse hoppings disabled (bottom-right panel). On the right, the magnified part of the diagram for the system with no transverse hopping is compared to the diagram for a fully connected array. Part (b) compares results for the fully connected reference array with results for the array with no transverse hopping (three chains not coupled by hopping).} %
\end{figure*}

\subsection*{Current bypassing obstacles - a single-site bottleneck}

    To verify this observation, we consider a transport obstacle that arises when the on-site energy of two of the three middle sites is detuned from the other sites. This modification creates a constriction forcing the current to flow through a single site in the middle column. In the regime of weak hopping, one might expect here results similar to those for a single quantum dot, \textit{i.e.}, much simpler diagrams \cite{FuechsleNN2012,WangCP2020,ZwanenburgRoMP2013}. However, this is not the case. In Fig.~\ref{fig:offsets-two}, we present the results for two non-equivalent ways to create a narrow constriction by detuning two sites in the middle column. The results are simple and additive in terms of the coexistence of single-site perturbations: we observe two pairs of lines being pushed outside the diagram, each independently corresponding to one of the detuned sites. The effective seven-site systems created in the extreme cases of high detuning still show diagrams qualitatively not much different from that for the reference array. All three slopes are still present in the diagram, suggesting that the conduction channels associated with the top and bottom rows are still active despite the current being forced through the single site in the middle column. Thus, we have another observation highlighting the crucial role of terminal sites with the terminal sites defining the slopes.

\subsection*{Impact of disabled hoppings}
    Another elementary disturbance that we can introduce into the system is disabling the hopping between a single pair of sites. We can do this in four non-equivalent ways for hoppings along the device, \textit{i.e.}, in the source-drain direction, and the corresponding results are shown in the first row of panels in Fig.~\ref{fig:hopping-cut}. Such a disturbance completely switches off the conduction channel associated with a given row of sites, which is observable by the disappearance of all lines with a given slope in the diagram. This outcome is understandable as each of these disabled hoppings involves a terminal site connecting the array to a lead. We have already seen that the terminal site is crucial for a given channel.
	
	The results look strikingly different when we disable transverse hoppings. The first three panels in the second row of Fig.~\ref{fig:hopping-cut} show that the changes due to removing transverse hopping are difficult to notice when compared to the reference array. Moreover, when all transverse hoppings are removed, as done for the last panel in the second row, the charge stability diagram is still very similar to the diagram for the fully connected, reference array. The only difference from the fully connected system is the closing of small anti-crossings in places magnified in the insets. In the last row of Fig.~\ref{fig:hopping-cut}, we compare Coulomb-blockade diagrams for the fully connected reference system and the one with the complete absence of transverse hoppings. Again, the results are difficult to distinguish from each other.
	
	Thus, the transport properties of the studied $3\times3$ array are strikingly similar to those of three $1\times3$ chains lying close enough for the electrons in neighboring chains to interact via the long-range Coulomb repulsion. This similarity suggests that transport is via conduction channels connecting the opposite lead contact points by the shortest available paths. In such a situation, the electrons do not exploit the transverse tunneling.

\section*{Conclusions}
    In conclusion, we have theoretically considered the transport properties of gated systems of $3\times 3$ phosphorus dopant arrays in silicon, which are now being used as analog quantum simulators and are promising for quantum information processing. Electrons in such a system are often strongly correlated because the Coulomb interaction can vastly exceed the hopping energy. Consequently, mean-field theories have to fail. For this reason, we have developed a methodology based on combining the exact diagonalization of the extended Hubbard model with the non-equilibrium Green's functions formalism. This methodology is a powerful tool for, \textit{e.g.}, transport calculations, previously typically done at a mean-field level. We have simulated actual devices recently demonstrated experimentally and presented the results for the stability and Coulomb blockade diagrams that are measured in experiments. We considered the ideal $3\times 3$ array with no disorder first. This allowed us to assess the impact of individual system parameters on the current flow. The simulated devices exhibited the same insulator-to-metal transition with increasing hopping that has been observed experimentally. Variations in the long-range Coulomb interaction and the exchange had little impact on the transport.
 
    Most importantly, we found that the transport shows significant resilience to disorder in the on-site energy with the charge-stability and Coulomb-blockade diagrams that characterize transport displaying qualitatively the same structure as seen in transport through a perfect array. By analyzing separately the effect of disorder from on-site energy fluctuations at individual sites, from single-site constrictions, and from vanishing hoppings, we found that the nature of the current flow is surprisingly simple. Conduction channels are defined by the terminal sites that connect to the source and drain, even when the electrons must pass around obstructions. The role of a terminal site is crucial, as its disorder may disable the whole channel. This simple picture of the transport emerges even when the electrons are strongly correlated. Significantly, this result does not depend on the array filling with electrons. Moreover, transport through the $3\times3$ array is very similar to the transport through a system with three closely located, Coulomb-coupled three-site chains with no interchain hopping.

    Simulating a realistic device, we also noticed that working above half-filling provided higher currents because low-energy traps in the array potential got filled up and screened at higher electron filling. Such an effect has also been observed experimentally \cite{WangNatureComm2022}. This observation may provide an additional method of optimizing the performance of future devices by working with the upper Hubbard band states.

\section*{Methods}\label{sec:methods}
We model the system using the extended Hubbard model defined by the Hamiltonian
\begin{align}\label{eq:hubbard}
H_{\mathrm{S}} = {}&{} -\sum_{\left\langle i,j\right\rangle}\sum_{\sigma} t_{ij}\,a_{i\sigma}^\dag a_{j\sigma}
+ \sum_{i} U_i \, n_{i\uparrow}n_{i\downarrow} \nonumber \\
{}&{} - \sum_{i} \left( \varepsilon_{i} - \mu_i \right) n_{i\sigma}
+ W\sum_{i,j}\sum_{\sigma,\sigma'}\frac{n_{i\sigma} n_{j\sigma'}}{\abs*{\bm{r}_i-\bm{r}_j}} \nonumber \\
{}&{} + J \sum_{\left\langle i,j\right\rangle}\sum_{\sigma} n_{i\sigma} n_{j\sigma}
- C\,W \sum_{i,j}\sum_{\sigma} \frac{n_{i\sigma} }{\abs*{\bm{r}_i-\bm{r}_j}},
\end{align}
where
$a_{i\sigma}^\dag$ ($a_{i\sigma}$) creates (annihilates) an electron with spin $\sigma$ at the site $i$ of the array positioned at $\bm{r}_i$,
$n_{i\sigma}=a_{i\sigma}^\dag a_{i\sigma}$ is the corresponding particle number,
$\left\langle i,j\right\rangle$ denotes nearest-neighbor pairs of sites,
$t_{ij}$ is the hopping for a given pair of sites,
$U_i$ is the on-site Coulomb repulsion energy,
$\varepsilon_{i}$ represents the single-particle electron energy at site $i$,
and $\mu_i$ is the local chemical potential.
Moreover, in the extended part of the Hamiltonian,
$W$ denotes the magnitude of the long-range electron-electron or electron-ion Coulomb interaction,
$J$ is the nearest-neighbor exchange interaction,
and $C = 1$ is the on-site charge of the positively charged sites.

For the simulation of the experimental device, we use site-dependent parameters from Ref.~\onlinecite{WangNatureComm2022}.
In the rest of the study where an ideal perfect array is the starting point, we take $t_{ij}=t$ and $U_i=U$.
The lever arms needed to calculate $\mu_i$ based on applied gate potentials are also calculated as in Ref.~\onlinecite{WangNatureComm2022} with the disorder removed for the ideal, perfect array.

We deal with a system where Coulomb repulsion is large ($U/t\sim50$ for the ideal device we model).
For this reason, we use exact diagonalization and take the resultant many-electron eigenstates as the starting point for Green's functions formalism to calculate the current in the system. 
Thus, we construct the matrix elements of the "bare" retarded Green's function $g^{\mathrm{R}}$ (which in our case already includes the impact of interactions) using the Lehmann (spectral) representation \cite{BruusBOOK2004}
\begin{equation}\label{eq:meth-lehmann}
\left[g^{\mathrm{R}}\left(E\right)\right]_{ij} = \frac{1}{Z}\sum_{m,n}\frac{\left(p_n+p_m\right)\matrixel[\big]{n}{a_i}{m} \matrixel[\big]{m}{a_j^\dag}{n}}{E-\left(E_m-E_n\right)+i0^+},
\end{equation}
where we have omitted spin indices,
$n$ and $m$ label the many-electron eigenstates of the system and run through all Fock-space sectors,
$p_n=\exp(-\beta E_n)$ is the occupation of state $n$,
$\beta=1/k_{\mathrm{B}}T$,
$Z=\sum_n p_n$ is the partition function,
and $E$ is the energy at which the function is evaluated.
The system is assumed to be in a stationary state and we work in the Fourier domain with respect to the time difference, $g(E)=\hbar^{-1}\int \mathrm{d}(t-t') g(t-t')\exp(-iE(t-t')/\hbar)$.
In addition to retarded functions, advanced Green's functions also appear, and the general relationship between them is $O^{\mathrm{A}} = ( O^{\mathrm{R}} )^\dag$.
We use a similar spectral representation to calculate the "bare" correlation functions $g^{\mathrm{<}}\left(E\right)$ and $g^{\mathrm{>}}\left(E\right)$ \cite{BruusBOOK2004}.

To attach the leads and thus study an open system, we use a standard embedding self-energy assuming semi-infinite one-dimensional leads \cite{DattaBOOK1995} and we include it via the Dyson equation \cite{BruusBOOK2004}, which yields the full retarded function $G^{\mathrm{R}}\left(E\right)$. Next, we apply the Keldysh quantum kinetic equation $ G^{</>}(E) = G^{\mathrm{R}}(E) \, \varSigma^{</>}(E) \, G^{\mathrm{A}}(E)$, to obtain full correlation functions $G^{</>}(E)$ \cite{HaugBOOK2008}. Here, $\varSigma^{</>}(E) = \sum_{l} \varSigma^{</>}_l(E) + \varSigma^{</>}_{\mathrm{int}}(E)$, where the first term contains standard lesser/greater self energies due to leads (index $l$ runs over the source and drain) calculated assuming they stay in thermal equilibrium \cite{DattaBOOK1995}, while the second term accounts for electron scattering in the system and can be calculated based on "bare" correlation functions $g^{\mathrm{</>}}(E)$,
\begin{multline}\label{eq:meth-g-int-less}
\varSigma^{</>}_{\mathrm{int}}(E) =\left[G^{\mathrm{R}}(E)\right]^{-1}\left[ \mathbb{I} - g^{\mathrm{R}}(E) \varSigma^{\mathrm{R}}(E) \right]^{-1} g^{\mathrm{</>}}(E) \\
\times \left[ \mathbb{I} - \varSigma^{\mathrm{A}}(E) g^{\mathrm{A}}(E) \right]^{-1}\left[G^{\mathrm{A}}(E)\right]^{-1}.
\end{multline}

To fully include the non-equilibrium distribution of occupations, we should do a self-consistent loop in which one step consists of recalculating $G^{\mathrm{R/A}}(E)$ by definition based on the difference of $G^{>}(E)$ and $G^{<}(E) $ (spectral function) and then applying Keldysh equation again. This turns out to be numerically challenging and will be addressed in a separate work. In this study, we have verified that for the studied low-bias conditions a single-step solution is sufficiently close to the converged one by comparing the anti-hermitian parts of $G^{\mathrm{R}}(E)$ before and after the step.

Finally, we follow Ref.~\onlinecite{MeirPRL1992} to calculate the terminal current
 \begin{align}\label{eq:meth-curr-terminal-final}
\left\langle j(E) \right\rangle = \, e \, \mathrm{Tr}\left[
\varSigma_{\mathrm{L}}^{<}(E) \, G^{>}(E) - \varSigma_{\mathrm{L}}^{>}(E) \, G^{<}(E) \right],
\end{align}
where subscript L denotes now either of the leads (currents on both terminals are identical). Integration over energy yields the current through the device.

\acknowledgments
    We acknowledge support from the National Science Centre (Poland) under Grant No. 2015/18/E/ST3/00583.
    We are grateful to Xiqiao Wang for the information on the experimental device and to Gabriela W\'ojtowicz, Marcin Mierzejewski, and Yaroslav Pavlyukh for their helpful comments.
    
%\section*{Data availability}\label{sec:data-avail}
%    The data supporting the findings of this study are available from the corresponding author upon reasonable request.

%\section*{Author contributions}
%    M.~G. formulated the model, performed analytical derivations, wrote the computer code, obtained the results and wrote the first version of the manuscript.
%    G.~W.~B. and M.~Z. initiated the project.
%    G.~W.~B. suggested relevant directions for the study.
%    M.~Z. supervised the work.
%    All the authors discussed the results and contributed to the manuscript.

%\section*{Competing interests}
%    The authors declare no competing interests

%apsrev4-2.bst 2019-01-14 (MD) hand-edited version of apsrev4-1.bst
%Control: key (0)
%Control: author (72) initials jnrlst
%Control: editor formatted (1) identically to author
%Control: production of article title (-1) disabled
%Control: page (0) single
%Control: year (1) truncated
%Control: production of eprint (0) enabled
%

\end{document}